# Heterogeneous component interactions: Sensors integration into multimedia applications

Christine Louberry
LIUPPA, Bayonne, France
Email: louberry@iutbayonne.univ-pau.fr

Philippe Roose and Marc Dalmau
LIUPPA, Bayonne, France
Email: {roose, dalmau}@iutbayonne.univ-pau.fr

*Abstract*—Resource-constrained embedded and mobile devices are becoming increasingly common. Since few years, some mobile and ubiquitous devices such as wireless sensor, able to be aware of their physical environment, appeared. Such devices enable proposing applications which adapt to user's need according the context evolution. It implies the collaboration of sensors and software components which differ on their nature and their communication mechanisms. This paper proposes a unified component model in order to easily design applications based on software components and sensors without taking care of their nature. Then it presents a state of the art of communication problems linked to heterogeneous components and proposes an interaction mechanism which ensures information exchanges between wireless sensors and software components.

## I. Introduction

Our work is interested in distributed applications based on software and physical components (sensors). Since few years, the technological developments in electronics and communication have allowed the arrival of mobile and ubiquitous devices providing several services. The growing demand for rich and customized services leads to the challenge of the realization of applications able to adapt themselves to the user's needs and to the real environment. The emergence of wireless mobile sensors able to process data in an autonomous way may allow proposing applications aware to their physical context and able to react according to the environment evolution. The characteristic of such applications is that they integrate strongly constrained devices. An example to illustrate that is a surveillance application (Fig. 1). We disseminate infrared sensors and camera sensors in an area we want to monitor. Infrared sensors can detect intrusions in the area. The detection of an intruder causes starting the nearest camera in order to obtain an image of the intruder. The collaboration of this camera with a video analysis software component enables to determine the probable trajectory of the intruder and to start the cameras located on this trajectory or to direct the cameras to obtain images of different angles of sight.

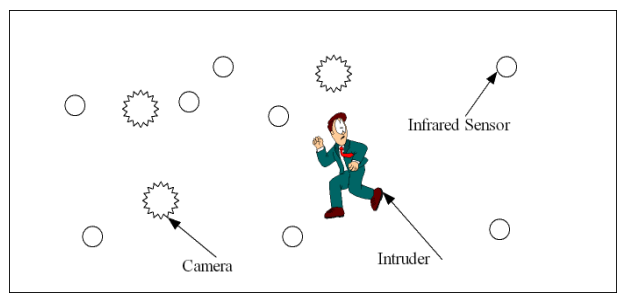

Figure 1. Example of an application based on software components and sensors

These applications need to integrate several types of components (software and hardware) but also several modes of communication (wired, wireless), several protocols (WiFi, ZigBee, Ethernet), and several mechanisms (method call, event, Mailbox, etc). Without the intervention of an intermediary, these components would not understand each other. To support interoperability, we chose to act at various levels. In this paper we focus on the level of application design. On the one hand we deal with the modeling of components by a unified component model then we deal with the modeling of interactions by a description of connections between components.

The reminder of this paper is organized as follows: In section II, we present related work on modeling sensors and interfaces between applications and sensors. Section III presents the wireless sensor model we use in our applications. In section IV we present a general view of our software component model called OSAGAIA and its various elements. We detail the contributions for the integration of sensors in the OSAGAIA model. Section V draws up a state of the art on communication mechanisms. In section VI, we discuss various approaches concerning data and protocol transformation facilitating communication between components. We conclude our paper and give the perspective of our research in section VII.

II. RELATED WORK

These last years, wireless sensor networks aroused the interest of research activities in computer science and electronic fields [1]. Most of them concentrate on energy consumption problems and on operating systems and network problems (routing, packet loss, connectivity). On the other hand, few activities are carried out around the problems of communication due to the heterogeneity of devices used. More precisely, preoccupations exist about the use of sensors to improve applications which until now are only run by software components.

For the moment, no model common to all sensors was proposed. Although there are standard communication protocols (WiFi, ZigBee, Bluetooth, etc.) and standard routing protocols for mobile networks (AODV [7], OLSR [12], etc.), there is no standard to model a sensor. In order to easily integrate sensors in applications and to propose a unified component model, we need a sensor model. We describe it in part III.

Moreover, due to low power and small memory, operating systems for sensors are low-level architectures and make application development non-trivial. To bridge the gap between applications layers (high and low), a new approach has emerged: middleware. In this paragraph, we present a survey of existing middleware, especially developed for sensor networks.

A classification of sensor oriented middleware according to their objectives can be found in [10]. The three main categories are: virtual machine based, database based and message-oriented middleware.

Virtual machine based middlewares allow developers to write applications in separate modules which are injected through the network. Then, the virtual machine interprets the modules. They run on the operating system of the sensor, that is to say they are embedded on sensors. Mate (TinyOS) [15] and Magnet (MagnetOS) [3] belong to this category.

In database based middlewares, the network is considered as a virtual database system. It offers a user-friendly interface to query the network and extract data. Cougar [5] uses a database approach to manage sensor network operation and TinyDB [16] uses queries to extract sensor data from a network using TinyOS.

Most of the time, sensors produce events. So, the most suitable communication model to this type of network is the asynchronous communication model. That's why message-oriented middlewares like Mires [21] propose a publish-subscribe mechanism. With this method, sensors only receive data which they are interested in.

A common point to these middleware is that they are used to facilitate the development of sensor-specific applications.

The researches mentioned above deal with applications embedded on sensors, dedicated to sensor networks but do not tackle the problem of collaboration between sensors and software components. However, in the future, sensors are intended to be used by applications which also integrate software components. Collaborations between these two categories of components will be necessary.

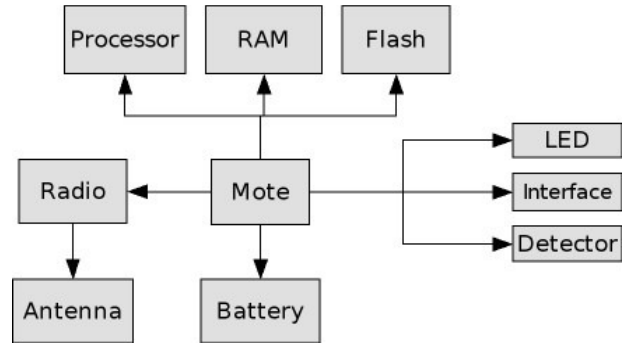

Figure 2. General architecture of a wireless sensor

The challenge is now to enable these components to communicate together in order to take advantage of the functionalities of sensors in applications and improve services.

III. WIRELESS SENSOR MODEL

The recent advances in microelectronics and wireless technologies allow developing small sized sensors endowed with processing capacities and wireless communication modes. Some of them allow even multimedia processing as sound and image (Cyclops [19]) thanks to embedded small cameras and microphones. This paragraph presents a brief state of the art and proposes a model for actual wireless sensors.

Sensors are generally composed of a core (mote) on which various components are attached (Fig. 2). A wireless sensor includes a processor, memory, a radio module, a battery and detectors [4]. It consists in three elements: an ID card, one or several functions and a communication module (Fig. 3). The ID card consists itself in four elements: a processor, a memory, a battery and an operating system. The communication module consists in a communication mode (for example event communication or client/server communication), and a communication protocol or a transmission type (WIFI or Bluetooth, etc). The communication module is endowed with a port allowing input/output of messages and events. For example, the ID card of a Crossbow MICA2 sensor consists of a processor Atmega128 at 4MHz, a 512 KB memory to store the measures, a system memory of 128 KB, a two AA battery and the TinyOS [11] operating system. It communicates by sending messages by radio.

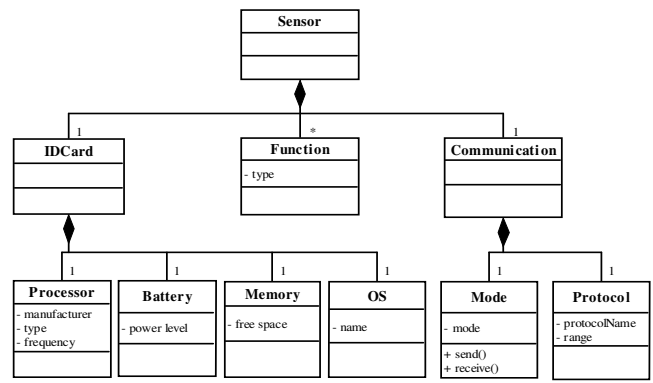

Figure 3. Sensor class diagram

A sensor can have several functions by integration of various detectors. It can measure outside temperature, atmospheric pressure, humidity, magnetic field, luminosity, displacement or capture pictures, sound, etc. Of course, when integrated into an application, this sensor will provide a precise service which will use one, some or all the functions of the sensor. The next paragraph describes this integration in a unified component model.

## IV. UNIFIED COMPONENT MODEL

In this paragraph, we proposed a unified component model to design applications without having to manage the software or hardware nature of components. This model is an adaptation, for wireless sensors, of the OSAGAIA model [8].

### A. OSAGAIA Model

The OSAGAIA model had been developed for distributed multimedia applications. It focuses on the problems of flow synchronization and of components dynamic connection/disconnection. It is made of two entities which handle that. The first one is the conduit that allows the transport of synchronous multimedia flows within the application. It can be distributed through the Internet. The second one is the Elementary Processor (EP). It is a container that provides a runtime environment for a Business Component (BC). The BC encapsulates a particular multimedia processing, i.e. the functional implementation. For instance, a video capture BC implements the necessary mechanism to provide this capture.

Inter-flow synchronization is known as temporal constraints between several flows (e.g. the sound and the image of a video), in opposition with intra-flow synchronization which concerns samples of the same flow. More precisely, these constraints are defined between the samples of each flow (e.g. one image corresponds to several sound samples in a video). So, it is necessary to identify the samples of each flow in a unique way in order to match them with samples of others flows. To do this, a time-stamp is associated to each set of samples on each flow at acquisition or creation time. We name this mechanism the flows time-stamping. The couple formed by a set of samples and a time-stamp is called a Temporal Unit (TU). A set of TUs corresponding to the same temporal interval from different synchronous flows is called a synchronous slice. Thus, a succession of synchronous slices constitutes a set of synchronous flows. They are bundled into a conduit in order to be transported.

Both entities of the model are connected by input/output ports. Ports are the means by which the multimedia flows pass from EP to Conduit, and conversely. Ports accept TUs as input or provide TUs as output. The port is the structural unit of connection between both entities of the model (connectable element). Output ports of one entity can only be connected with input ports of others ones.

BC implements a particular media processing (functional implementation). The BC needs to be executed in a container named EP. The BC is data driven, that means its processing is linked to incoming data.

The EP is a container for the BC. It supports non-functional properties for a correct execution of the BC (functional properties) and of the whole application. The EP is composed of an Input Unit (IU), an Output Unit (OU) and a Control Unit (CU) as shown in figure 4. The EP is supervised by the platform (add, remove or replace EPs). The EP has input/output ports for each multimedia flow entering or outgoing. These ports allow its connection to conduits. Each port is linked respectively to the IU or the OU. These units are interfaces between BC and multimedia flows. They contain methods used by the BC in order to read (respectively write) in the input (respectively output) ports. The CU manages all the elements of the EP. This unit communicates with events and specific methods. For instance, the BC behavior is controlled by the CU through its methods init(), start() and stop(). CU also manages the data circulation within the EP. Particularly it ensures that incoming flows which are not processed by the BC cross the EP without loosing the synchronization between them and with flows which are processed. A prototype is available to the following URL: http://www.iutbayonne.univ-pau.fr/~roose/V2/korronteaSimulator/KorronteaSimulator.zip.

Sensors are particular Business Component. We extend OSAGAIA model in a unified model to integrate them in applications.

### B. Unified Model

Sensors are able to produce several kinds of data flows. To process information, they communicate with software components able to achieve the specific processing of this information. To integrate sensors among software components, we have to propose a unified component model. We propose to integrate a sensor into an Elementary Processor (EP) of the OSAGAIA model. The EP encapsulates the sensor as it would do for a software component.

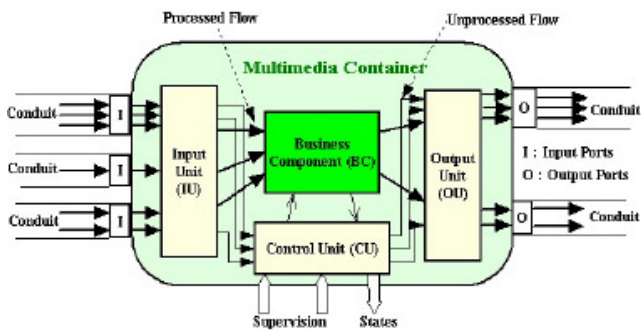

Figure 4. Internal architecture of an Elementary Processor

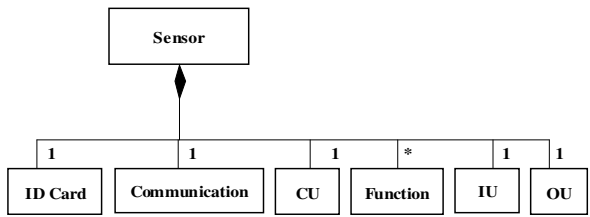

Figure 5.  Class diagram of an Elementary Processor

Using the OSAGAIA model, the interconnection of components is done using an Input/Output Unit (*IU, OU*). The execution platform supervises the Business Component (*BC*) thanks to a Control Unit (*CU)* located into the container (*the Elementary Processor - EP*). According to OSAGAIA, the Business Component (BC) is used to process multimedia flows. A flow enters the Communication Unit via the Input Unit of the Elementary Processor (EP) and get out through the Output Unit. These units are supervised by the Control Unit of the EP. In order to inter-connect and to manage the sensor, we add a CU, a IU and a OU to it (Fig 5). The CU allows to send commands to the sensor and to the IU and OU and to get back their state. The CU is able to evaluate the state of the memory and the battery of the sensor in order to inform in real-time the supervision platform about the available space or the battery level. It can also communicate with the sensor OS in order to supervise it.

In the OSAGAIA model, the supervision platform is distributed on all sites. Because of memory size restrictions and compute power limits it is not possible to locate a part of this platform on each sensor as it is traditionally done on each computer. That is why we choose to externalize the CU associated to the sensor to the nearest site able to support the platform. This externalization is not reflected in the UML diagram because, at a structural level, the Control Unit is part of the Elementary Processor. Actually, the role of the CU is to ensure the link between the component and the platform.

Using this process, the model obtained (Fig. 5) matches the model of the Elementary Processor in the OSAGAIA model.

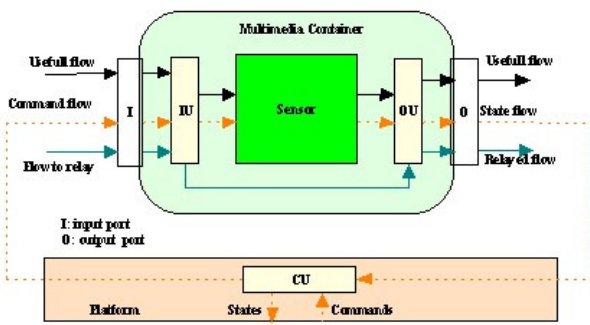

Figure 6.  Integrating a sensor into an OSAGAIA EP

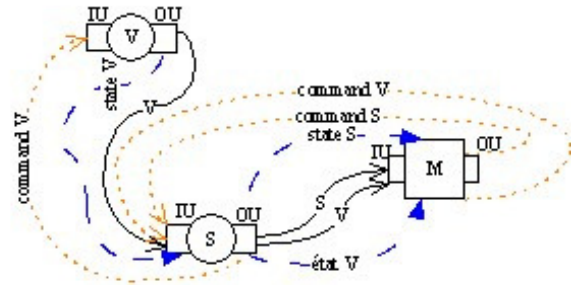

V : video sensor
S : sound sensor
M : mixing software component on a non-mobile terminal

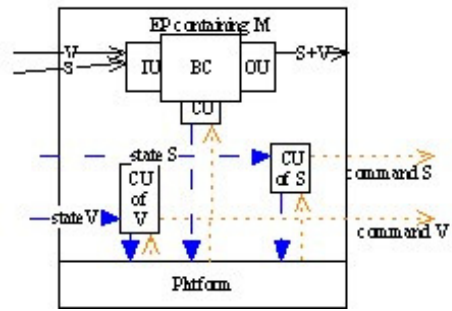

Figure 7.  Example of flows exchanges into an application composed of mobile and non-mobile components

However, a sensor communicate with its radio (wireless network card, etc). It is its only interaction point with other components. Consequently all information exchange will be done using the Input/Output radio device (Fig. 6). So, we need to distinguish data and control/state flows in order to re-orientate them according to their nature towards the corresponding entity.

That is why we use a data flow model including the information of course (data, command) but also an identifier allowing knowing if this flow is:
- a data flow;
- a state flow;
- a command flow.

The figure 7 shows an application composed of a mobile video sensor (V), a mobile sound sensor (S) and a mixing software component (M) located on a non-mobile terminal. On the below part of the schema, a zoom on this non-mobile terminal shows the local part of the platform, the Control Units of the two sensors (V, S) and the Elementary Processor containing the Business Component M. The sensor V sends a video flows to M, but because M is too far, S has to play a relay role. S receives this flow, identifies it as a flow to relay and communicates it to its Output Unit in order to transmit it to M. S also sends its own produced flow to M. M reads the two flows received into in Input Unit, identify them as data flows and communicates them to its Business Component. When the platform needs to send a command to V, it sends it to S which relay it to V.  This is the same when S and V send state flows to the platform. However, in order to not overload sensors, the platform is distributed on all non-mobile stations; because of the mobility of sensors, the Control Units of sensors can be moved from one fixed station to another in order to directly reach the sensor if possible. This is part of the quality of service management that the platform normally does.

This process allows managing both sensors and software components in a unique way thanks to the generic model.

Now, there is a generic model to manage sensors and software components, we have to propose a mechanism facilitating communication between these components. Next paragraph presents such a mechanism.

## V. STATE OF THE ART ABOUT COMMUNICATION MECHANISMS

This paragraph presents a state of the art of communication mechanisms that software components and sensors can both use.

### A. Communication in software components

In the software engineering literature, we can notice that the software components communication mechanisms most used are event-based, method call-based, stream-based, client-server-based and message-based communication. The first one is generally used to report property changes of a component to others. Method call is the traditional communication mechanism of software components. Calls can be local or distant (RPC). The stream-based mechanism is often used to transfer multimedia data. The next part shows that all mechanisms cannot be used by sensors due to their operating system.

### B. Communication in wireless sensors

The most popular wireless sensors are Crossbow MICA2 and Java Sun Spot. The first ones use TinyOS operating system [11]. TinyOS proposes a communication mechanism by messages and uses its own messages format. This format looks like a network packet [14]. It encloses the address of destination, the length of the message and the data field. The data field can contain many kinds of data (measurements, video, sound) but also others data structures like commands. The second ones use Squawk java virtual machine. Squawk runs without any operating system. It proposes a message-based communication mechanism too but looking at the Java Sun Spot API, we can notice that Sun Spot can also use a client-server-based communication mechanism and a stream-based one. In reality, usage of the radio link reduces these possibilities to only one: the message-based mechanism.

### C. Interaction modelling

Modeling interactions is a recurrent challenge in software engineering. One way to describe interactions is to use an architecture description language (ADL) [2]. Such languages introduce the concept of connector. In [18], authors draw up a taxonomy of connectors. They classify connectors into four service categories: communication, coordination, conversion and facilitation connectors. We focus on conversion connectors and more precisely, on adaptors. Adaptors are a kind of connectors which provide facilities to components to interoperate although they have not been designed for.

Several researches were made in order to enable heterogeneous software components to interoperate. Indeed, due to the several models proposed and to the reuse of components preoccupation, applications based on software components come up with the problem of technical and semantic heterogeneity. Most of these works deal with semantic interoperability and propose solutions to bridge the gap between incompatible interfaces' signatures.

In this article, we are interested in adaptors and technical interoperability. Adaptors also called wrappers are piece of code linking two components that normally have incompatible interfaces. Examples of this kind of connectors are adaptors of Yellin and Strom [23] or wrappers of Spitznagel and Garlan [22]. They propose to construct adaptors using finite state machines (FSM). These adaptors notice differences between the FSM of communication protocols of two components and provide some code hiding these differences and allowing components to interact. In [23], authors define wrappers as new code that moderate the behavior (data format, protocol of interaction, etc) of components without modifying it. Their work focuses on wrappers that affect the communication between components. They specify connector wrappers as protocol transformations able to redirect, replay, insert and discard particular events.

As we are interested in protocol translation, we want to define connectors allowing to link two components using different communication mechanisms. For example, we want a sensor using message-based communication to interoperate with a software component using method call-based communication. Defining such connectors requires knowing all the types of interactions we can encounter in an application. The next paragraph presents a list of the communication mechanisms that components (software and sensors) can use.

## VI. COMMUNICATION PROTOCOL

Interactions between heterogeneous components are recurrent problems in software development. Our applications make collaborate two kinds of components: software components and wireless sensor. Wireless sensors and software components differ on several points. They have different nature, hardware and software, they use different communication mechanisms, etc. They need to communicate in order to ensure service collaboration. We first propose solutions which deal with interactions between sensors and software components in a general way. Then we consider communication in the Unified Model. As an EP is a container for a sensor, we have to link them so that they can exchange information.

### A. Heterogeneous component interactions

Software components and sensors do not use the same communication mechanism. First one uses method calls,

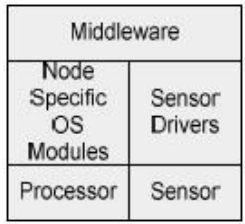

Figure 8. Structure of a sensor software [4]

whereas the other one uses messages broadcast. We have to provide a mechanism that acts as a link between such elements.

A first approach consists in introducing an interaction transformation process into the input and output units (IU and OU). The first possibility is to add an interaction transformer to the OU. When a component sends data to another, the OU transforms it in the appropriated format for the addressee. This mechanism implies that the component knows the addressee's type. It is not true because only the supervision platform knows the components' type. Secondly, we can add a data transformation process to the IU. When a component receives a data, the UI identifies it in order to apply the appropriate transformation. This method implies that the entry port of the destination component can accept data in any form. It also implies to know the data structures and interaction mechanisms of all the components of the network. In the case of sensors, this method is not applicable due to their small memory. Moreover, each component must know all the possible transformations. It means that when a new transformation is introduced into the application, all components have to be updated. That will be difficult to deploy on a real scale.

A second approach consists in using a middleware. A description of the characteristics required by a middleware for sensor networks can be found in [4]:
- scalable: the application is reduced to essential components and data types.
- generic: interfaces must be generic to minimize customization for other applications.
- adaptive: able to add/remove components during runtime.
- reflective: able to change the behavior of components instead of changing themselves.

The authors propose a concept of a software-architecture for wireless sensor networks which separates software from hardware and divides the software into three functional blocks (Fig. 8). The Node-specific Operating System handles device management, for example boot up, memory management, etc. Sensor drivers groups hardware drivers, e.g., timer, radio. Middleware then organizes the collaboration of nodes (collaboration of services). With this architecture, sensors integrate a distributed middleware which is the only way to contact them in order to simplify the development of services for sensor networks. The authors of [9] propose a middleware pattern for sensor networks in order to handle the heterogeneity in sensor applications. It combines services proposed by existing middlewares for sensor networks. Services are divided in three categories: Application layer, Data management layer and Network service management layer. They are implemented in separate components in order to make it possible to replace them. Applications indicate their needs to the data management layer which gathers the needed data by interrogating sensors. Reusing readings enables to save energy but is not suitable to realize real-time applications. Moreover, middleware is suitable to reconfiguration but not to data transformation. Because components do not care about their neighbors, the middleware would have to request the supervision platform before each sending in order to transform interaction mechanism. With sensors, it would generate too many transmissions.

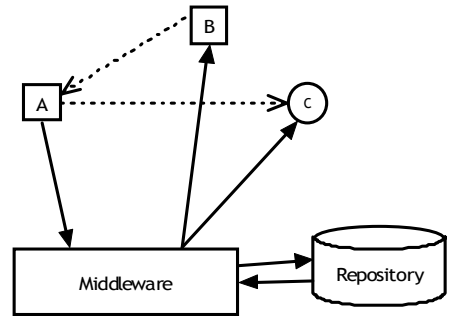

Figure 9. Example of a centralized middleware and its repository

In our model, we already integrate input and output units in a sensor. Adding a middleware could harm the operation of the sensor due to its low power and its small memory.

Instead of a distributed middleware, we can use a centralized middleware with a repository which contains all the data type transformations. Figure 9 shows an application composed of two software components A and B and one sensor C. Instead of sending two messages in two different formats to B and C, A sends its message to the middleware which transforms and sends it to B and C with the appropriate format. However, the use of such a middleware increases networks transfers and add delays because of transaction time with the repository.

A third approach consists in using OSAGAIA software components [8]. We can define some Business Components (TC) which provide a conversion processing specific to each kind of components (Fig. 10). Each component is associated to the conversion component specific to it. This method limits delay because it only induces some processing time whereas middleware method induces network transfer time. Another advantage is the preservation of the synchronization. Indeed, the conversion component is a component of our model and consequently contains the properties to keep the synchronization.

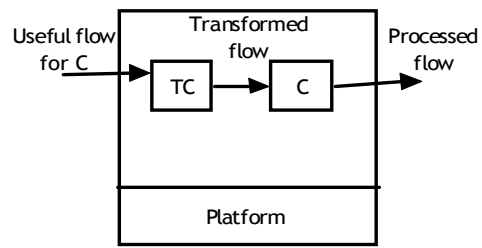

Figure 10. Example of conversion component

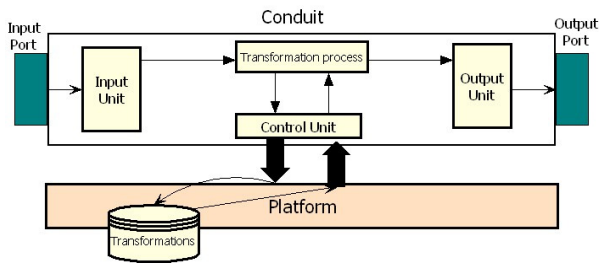

Figure 11.  Transformation in the Conduit

The disadvantage is when we reconfigure the application. We have to change the components per pair: the component and its conversion component. It implies of being aware of functional dependencies between components as in [13].

A fourth approach consists in using the Control Unit (CU) of the Conduit in our model (Fig. 11). In the OSAGAIA model, all data streams are transported by Conduits. The Conduit contains synchronization properties that allow keeping the synchronization during data transport. The purpose is to implement the CU in order to know all possible data transformations in the network. There is no more network delay, only processing time due to the data transformation. This solution is the most suited to our applications. Conduit is a kind of middleware independent of business components. It ensures communication transparency and in case of sensors, does not introduce additional transmissions and processes what preserves its resources.

Obviously, all the methods described in this paragraph require knowing all the data types which will be used in the network. They also imply that the application must know the composition of the network permanently in order to give messages to the appropriate transformation component according to the destination.

This paragraph summarizes general solutions about heterogeneous component interactions. Now we focus on one kind of interaction: communication in the Unified Model.

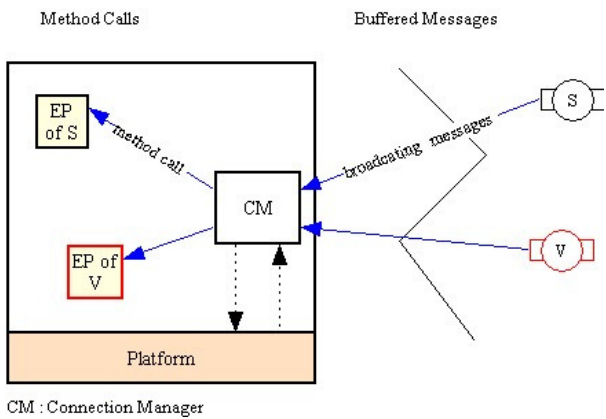

Figure 12.  Connection between EPs and sensors

### B. Interactions in Unified Model

All the solutions presented before are available for a massively heterogeneous network with many types of components. In our model, we consider that all business components, whatever their nature is, are contained in an EP. However, the communication mechanisms of software components and sensors are different. Most of the time, software components exchange data according to local or remote procedure calls. It implies that components have public methods invoked by other components. However, sensors cannot use procedure call to communicate with other components using their radio link.

When an EP encapsulates a software business component, the two entities are logically located on the same platform (or base station) and communicate via method calls. When an EP encapsulates a sensor, in order to not overload CPU and memory capacities, we decide to export EP functionalities on the nearest base station from the sensor. As we said in the previous paragraph, contrary to business component, sensors communicate broadcasting messages to the EP. Indeed because the two entities are on two distinct platforms, we have to define a connection mechanism between the EP and the sensor it encapsulates. This connection must include a communication mechanism adaptor from method call to broadcasting (mailbox) and inversely (Fig. 12).

In [6] and [17], authors distinguish two kinds of communication abstraction: connector and medium. They define a connector as an abstract architectural element. It specifies the reification of an interaction, communication or coordination system of an application. It provides some extra-functional interaction mechanisms independent of the application. A connector provides generic interfaces which are adapted to the specifications of the linked components' interfaces. This mechanism ensures the transparency of the communication.

The medium reify a communication or interaction abstraction. It is a software component which offers a communication service. A medium provides explicit interfaces with methods that components can invoke directly. Unlike a connector, a medium is dependant of other components; their implementation must integrate the use of the communication service.

Our goal is to ensure collaboration between heterogeneous components and to provide a connection mechanism transparent to the EP. Thus an EP can integrate a business component or sensor without taking care about the connection. Connectors seem to be the most adapted way to reach our objective.

The next paragraph describes the interaction abstraction we choose to use for this connection.

### C. Communication mechanism connector

The sensors we are interested in are smart sensors like Sun SPOTs (Fig. 13). Indeed, the applications we implement are dynamically reconfigured according to the environment evolution. Crossbow sensors (Motes) use the TinyOS operating system. To run an application with TinyOS, you first have to create this application as a TinyOS module, then to create an image of the operating system including this new module and finally to load this image on the sensor. This process is too heavy and is not suitable to dynamically reconfigurable real-time applications. Loading a new operating system image at each reconfiguration would spend too much energy and time. Sun SPOTs sensors do not use an operating system but a Java virtual machine which is more suitable to reconfiguration. The main characteristics of Sun SPOTs are:
- Microcontroller 16Mhz
- 512Kb RAM and 4Mb Flash memory
- Wireless communication 802.15 ZigBee compliant
- Squawk Java virtual machine J2ME CLDC 1.1 compliant running without any operating system [20].

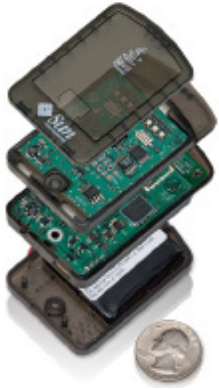

Figure 13. Sun SPOT wireless sensor, sensor board on top, processor and radio in the middle and battery board on the bottom

As the definition of a connector described in [17], we define a connector to link an EP and its sensor with its property, its plugs and its protocol.

The property of the communication mechanism connector is to ensure the adaptation of communication mechanisms of the EP and the sensor so that they can interact.

The connector has two plugs. The first plug, called Left plug, receives the requests of the EP and transmits the answers from the sensor. Its interface has the same public method than the EP. Thus, Left plug communicates with the EP by method calls. The second plug, called Right plug, waits for messages coming from the sensor. It transforms the method calls in comprehensible message by the sensor and conversely. The Right plug communicates with the sensor by messages broadcast.

Interactions between components imply to follow some rules in order to organize the communication. Theses rules are defined with a protocol. Because of the unreliability of the sensors (connectivity, battery), we decide that the best way to offer a suitable quality of service to the user in case of material breakdown is to propose an asynchronous communications protocol. Thus, if for an unspecified reason, the sensor breaks down whereas its EP requests it, the EP would not remain blocked waiting for an answer, blocking a part of the application. Another procedure allowing discovering the devices of the network would be used to inform the EP of the absence of the sensor. That causes a quality of service event which is caught by the supervision platform. The platform moves the EP in order to allow it reaching the sensor. If it is not possible, the platform chooses a sensor to ensure a relay function between the EP and the too far sensor. Figure 14 represents the interaction diagram of an EP communicating with its sensor. The Right plug creates a message each time it receives a corresponding EP method call from the Left plug. This creation process is based on the following scheme:

Message name = Method name;
Message property = {Method parameters};
Write message in output port;

The connector is located on the same base station as the EP. Thus, exchanges with the EPs and adaptation processes are carried out locally to the base station, only the already transformed information is transmitted to sensors.

Providing a connector to link sensors and EPs does not entirely resolve the communication problem. We now face with mobility problems. When the sensor moves, appears or breaks down, we have to transfer, add or remove the EP and the connector to the nearest base-station. It implies to provide a process which will be aware of network composition.

## VII. CONCLUSION AND FUTURE WORK

Sensors become more and more present around us. They have now processing capacities, a relatively important memory and can do measures and capture sound or picture. Our objective is to use them to improve multimedia applications by adding services linked to the physical context.

In order to design such applications easily, we propose a unified component model allowing the developer not to take care of the type (hard/soft) of entities. In this paper, we focused on the OSAGAIA model and show how to extend it to sensors. However, we had to take into account the low capacity and the mobility of sensors. A prototype implemented with JavaBeans is available and allows simulating the deployment of sensors/software components and their mobility.

This original model allows designing applications using inter-connections of hardware and software components without any particular adaptation of the components involved. The platform is able to supervise these components and can re-organize the circulation of data flows to improve the QoS of the application. It receives states from each of them in order to know how the application runs and sends command to the components in order to drive the execution.

Within sight of the various solutions of data management described in part VI, we can see that there is a real need with regard to data transformation and data management. The majority of the solutions deal with applications specific to sensor networks. Few ones are interested in the problems of integration of the sensors in existing applications. The approach we propose is interested in the problem of components heterogeneity in applications which mix software and hardware components.

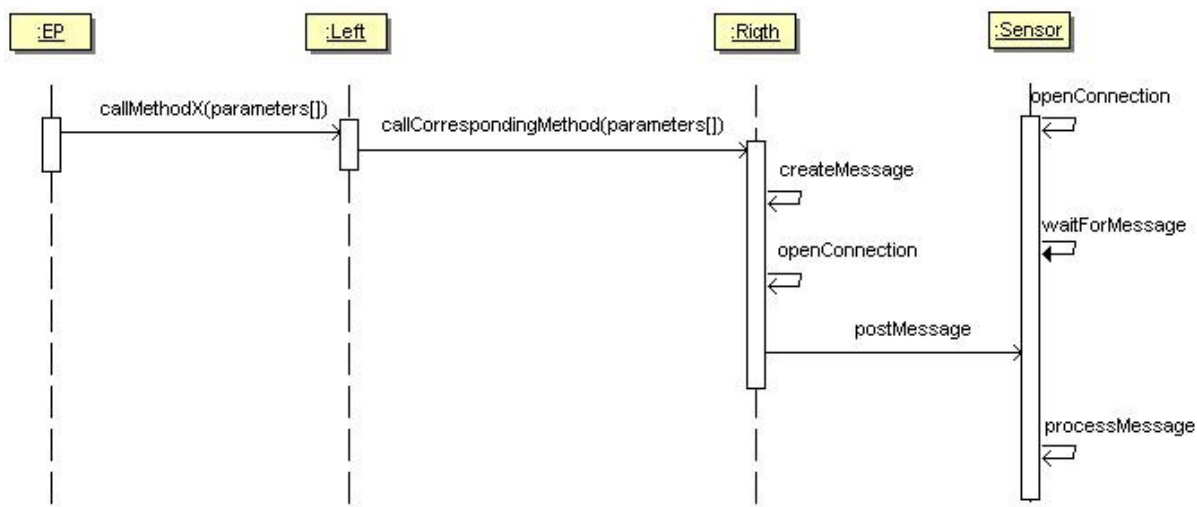

Figure 14. Asynchronous communication protocol

Future works will be in the discovery of devices in the network in order to manage connection between EP and sensors. The most popular service discovery protocols are UPnP and Jini. UPnP (Universal Plug and Play) is an industry standard to allow devices to be automatically discovered and added into a network in an easy-to-use way. UPnP is based on TCP/UDP and HTTP protocols. Jini Technology proposes a service discovery protocol for adhoc network. It is a Java-specific middleware that can only be used by client able to interpret Java bytecodes. Due to their characteristics, Sun SPOTs can integrate one of these service discovery protocols.

Some tests were implemented on Felix OSGi framework. We developed a group of connectors according to the specification we propose, mapping a method call from the EP to a message the sensor can understand (http://www.iutbayonne.univ-pau.fr/~louberry/Pages/recherche.html). Next simulations and tests will be dedicated to the integration of service discovery protocol in our application in order to improve the communication.

ACKNOWLEDGMENT

This work is supported by the ANR / CNRS, 2006-2009.

**Christine Louberry** is a Ph.D. student in the department of Computer Science at the University of Pau, France. She obtains her Master Degree in Computer Science from the University of Pau, France, in 2006. Her research interests include sensor networks, component model and software architecture for distributed multimedia applications, quality of service.

**Philippe Roose** is an Assistant Professor in the department of Computer Science at the University of Pau, France. He is responsible of the TCAP project - Video flows transportation on sensor networks for on demand supervision. His research interests include wireless sensors, software architectures for distributed multimedia applications, software components, quality of service, dynamic reconfiguration, COTS, distributed software platform, information system for multimedia applications.

**Marc Dalmau** is an Assistant Professor in the department of Computer Science at the University of Pau, France. He is a member of the TCAP project. His research interests include wireless sensors, software architectures for distributed multimedia applications, software components, quality of service, dynamic reconfiguration, distributed software platform, information system for multimedia applications.